# Generating surrogate data for time series with several simultaneously measured variables

Dean Prichard*
*Department of Physics, University of Alaska, Fairbanks, AK 99775*

James Theiler
*Santa Fe Institute, 1660 Old Pecos Trail, Santa Fe, NM 87505;* and
*Center for Nonlinear Studies and Theoretical Division, Los Alamos National Laboratory, Los Alamos, NM 87545*
(March 30, 1994)

We propose an extension to multivariate time series of the phase-randomized Fourier-transform algorithm for generating surrogate data. Such surrogate data sets must mimic not only the autocorrelations of each of the variables in the original data set, they must mimic the cross-correlations *between* all the variables as well. The method is applied both to a simulated example (the three components of the Lorenz equations) and to data from a multichannel electroencephalogram.

## I. INTRODUCTION

A number of measures have been developed for quantifying deterministic low-dimensional chaotic behavior as manifested in a time series; these include estimates of the dimension of the strange attractor [1], of the Lyapunov exponent(s) [2], and of nonlinear prediction error [3]. Computing these quantities can be problematic, however, and values can vary markedly from one algorithm to the next. Furthermore, nonchaotic and even linear stochastic processes can generate time series data which these algorithms may incorrectly characterize as low-dimensional [4,5]. For this reason, a number of authors [6–8] have advocated a direct comparison of the measured data set with computer generated "surrogate" data sets that have the same linear correlations as the original.

The basic idea is to compute the *nonlinear* statistic of interest for the original data set and for each of an ensemble of surrogate data sets. If the computed statistic for the original is significantly different from the values obtained for the surrogate sets, one can infer that the data were not generated by a linear process; otherwise, there is no reason to reject the notion that a linear model fully explains the data. Surrogate data can provide a formal statistical test of the null hypothesis that the data are linear, or an informal "sanity check" on whether an estimated dimension, say, is anything more than an artifact of linear autocorrelation.

For univariate time series, two approaches have been suggested for generating surrogate data consistent with the null hypothesis of linearly correlated gaussian noise. One approach is to fit an explicit linear model to the data (*e.g.* an autoregressive moving-average, or ARMA, model [9]) and then to iterate the model to generate the data [6]. A second approach is to Fourier transform (FT) the data set, randomize the phases, and then invert the transform [7,8]. It is beyond the scope of this Letter to discuss the practical and theoretical differences between the two approaches; we will focus on the FT method because it is the more straightforward of the two to implement.

Though much of the work on nonlinear time series analysis has focused on univariate data, one often has available several simultaneous measurements of a system, either of different aspects (pressure and temperature, say), or at different spatial locations. For instance, it is conventional to simultaneously measure electroencephalogram (EEG) signals from various places on the scalp, and a number of authors have used this multivariate data for dimension estimation [10,11]. As with univariate time series, one would like to use surrogate data to assess the role of linear correlations in contributing to the relatively low dimensions that were reported in these studies.

In the next section, we describe an algorithm for generating multivariate surrogate data that corresponds to the null hypothesis of linearly correlated gaussian noise. In Section III, we apply this algorithm both to a simulated and to a real multivariate data set. We find in both cases that the evidence for nonlinear structure can be (but is not necessarily) stronger for the multivariate data set than for any of the individual variables. More detailed investigations will be reported elsewhere [12].

## II. SURROGATE TIME SERIES

As a brief review, and to introduce the notation, we will first describe how to generate univariate phase-randomized Fourier-transform surrogate data. Given a time series, $x(t)$, of $N$ values taken at regular intervals of time $t = t_0, t_1, \ldots, t_{N-1} = 0, \Delta t, \ldots, (N-1)\Delta t$, apply $\mathcal{F}$, the discrete Fourier transform operator, to obtain

$$X(f) = \mathcal{F}\{x(t)\} = \sum_{n=0}^{N-1} x(t_n) e^{2\pi i f n \Delta t}. \quad (1)$$

Further, write this complex valued Fourier transform as: $X(f) = A(f)e^{i\phi(f)}$, where $A(f)$ is the amplitude and $\phi(f)$ is the phase. $X(f)$ is evaluated at the discrete frequencies $f = -N\Delta f/2, \ldots, -\Delta f, 0, \Delta f, \ldots, N\Delta f/2$, where $\Delta f = 1/(N\Delta t)$.

A "phase-randomized" Fourier transform $\tilde{X}(f)$ is made by rotating the phase $\phi$ at each frequency $f$ by an independent [13] random variable $\varphi$ which is chosen uniformly in the range $[0, 2\pi)$. That is,

$$\tilde{X}(f) = A(f)e^{i[\phi(f) + \varphi(f)]}, \quad (2)$$

and from this, the surrogate time series is given by the inverse Fourier transform:

$$\tilde{x}(t) = \mathcal{F}^{-1}\{\tilde{X}(f)\} = \mathcal{F}^{-1}\{X(f)e^{i\varphi(f)}\}. \quad (3)$$

By construction, $\tilde{x}(t)$ will have the same power spectrum as the original data set $x(t)$, and by the Weiner-Khintchine theorem the same autocorrelation function [15].

For multivariate time series, we not only want our surrogate data generator to reproduce the linear properties of each of the time series, but also any linear correlations between them. Suppose we have $m$ simultaneously measured variables, $x_1(t), x_2(t), \ldots, x_m(t)$ with zero mean and unit variance, and let $X_1(f), X_2(f), \ldots, X_m(f)$ denote their respective Fourier transforms. The cross-correlation between the $j^{\text{th}}$ and $k^{\text{th}}$ variables is given by: $C_{jk}(\tau) = \langle x_j(t) x_k(t - \tau) \rangle$. For a linear gaussian multidimensional process, all of the information about the process is given by these cross-correlations. By an extension of the Weiner-Khintchine theorem, the Fourier transform of the cross-correlation function is the cross-spectrum:

$$X_j^*(f) X_k(f) = A_j(f) A_k(f) e^{i[\phi_k(f) - \phi_j(f)]}, \quad (4)$$

where again $A(f)$ is the Fourier amplitude, and $\phi(f)$ is the phase angle. To preserve all the linear auto- and cross-correlations, we need to fix $X_j^*(f) X_k(f)$ for all pairs $j, k$. Since Eq. (4) only involves differences of phases, this is readily achieved by adding the *same* random sequence $\varphi(f)$ to $\phi_j(f)$ for all $j$. That is,

$$\tilde{x}_j(t) = \mathcal{F}^{-1}\{X_j(f) e^{i\varphi(f)}\}, \quad (5)$$

where $\varphi(f)$ is the same for all $j$.

## III. APPLICATION TO MULTIVARIATE DATA

### A. Lorenz equations

As an example, we compare multivariate and univariate embeddings for $N = 512$ points from the Lorenz equations [16] (with parameters $\sigma = 16$, $\beta = 4$ and $R = 45.92$). The sampling time $\Delta t$ is varied from 0.02 to 1.00 in increments of 0.02. For each choice of sampling time we create time series of the $x$, $y$, and $z$ components, and make 39 univariate surrogates of each component individually, as well as 39 multivariate surrogate data sets. We account for the nongaussian amplitude distribution by using the amplitude adjusting algorithm described in Ref. [8] for each component.

For our discriminating statistic, we use the Takens best estimator of correlation dimension [17]

$$Q = D_{\text{Takens}} = \frac{C(r_o)}{\int_0^{r_o} (C(r)/r) dr} \quad (6)$$

where $r_o$ is an upper cutoff, and $C(r)$ is the correlation integral

$$C(r) = \frac{2}{N^2} \sum_{k=W}^{N-1} \sum_{j=0}^{N-1-k} \Theta(r - \|\vec{x}(t_{j+k}) - \vec{x}(t_j)\|). \quad (7)$$

Here, $\Theta$ is the Heaviside function, $\|\cdot\|$ is the maximum norm, and $W$ is a constant, the order of a few autocorrelation times, which is used to remove autocorrelative effects [4]. $\vec{x}$ can either be a multivariate signal, or a time delay embedding [18]: $\vec{x}(t) = [x(t), x(t - \tau), \ldots, x(t - (m-1)\tau)]$.

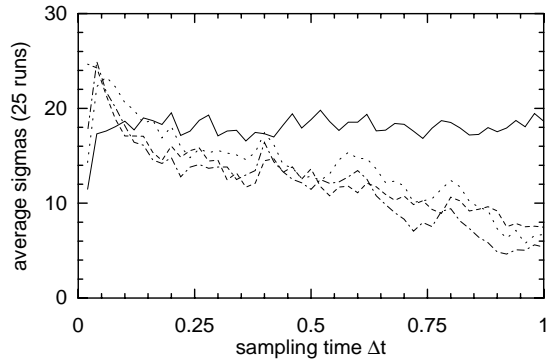

FIG. 1. Average significance (measured in "sigmas") of the finding of nonlinearity in a Lorenz time series with $N = 512$ points using the Takens dimension estimator with $r_o = 6.5$ as a function of the sampling time $\Delta t$. The solid curve is for the multivariate embedding $(x, y, z)$, while the dashed line is for a univariate embedding of the $x$ component, the dot-dashed line is for the $y$ component, and dotted line is for the $z$ component.

The Takens estimator with upper cutoff $r_o = 6.5$ (roughly half the standard deviation of the series) and $W = 5$ is computed for each of the $x$, $y$, and $z$ components as well as for their surrogates using a time delay $\tau$ equal to the sampling time and embedding dimension $m = 3$. We also calculate the Takens estimator for the multivariate embedding (simultaneous $x$, $y$, and $z$) and its multivariate surrogates. For each choice of sampling time and embedding ($x$, $y$, $z$, and multivariate) we use the following rough measure of significance: $S = |Q - \langle Q_{\mathrm{surr}} \rangle|/\sigma_{\mathrm{surr}}$ where $Q$ is the Takens estimator for the original data set, $\langle Q_{\mathrm{surr}} \rangle$ is the mean value of the statistic for the surrogates, and $\sigma_{\mathrm{surr}}$ is the standard deviation of the of the statistic for the surrogates. The units of $S$ are commonly called "sigmas". The whole process is then repeated 25 times using new sequences of $x$, $y$, and $z$ from the Lorenz equations and the significance is averaged over all 25 runs. In Fig. 1 we show the average significance as a function of sampling time for the $x$, $y$, $z$, and multivariate embeddings. The figure shows that for sampling times shorter than the mutual information time ($\Delta t \approx 0.11$), it is easier to detect nonlinearity using the univariate embeddings, while for longer sampling times multivariate is better.

### B. Human electroencephalogram (EEG)

As a second example, we apply the multivariate surrogate data method to 16-channel EEG data, recorded for two minutes at 128 Hz from a 20 year old female volunteer in a relaxed state with eyes closed. These data were generously supplied by Milan Paluš, and are more fully described in Ref. [11]. Paluš *et al.* [11] have also analyzed this data set and they compute a correlation dimension of 5.8, though they argue that this number should not be interpreted "as a dimension of a hypothetical strange attractor," but instead as a measure of the average "complexity" of the signal. Complexity, of course, is a difficult concept to quantify, but an estimated correlation dimension can still provide a discriminating statistic in tests for nonlinear structure. In this letter, we describe results for the first 8192 points. The same analysis was applied to the last 8192 points with essentially the same results. Before making the multivariate surrogate data sets, we first filter the data with a simple notch filter in the frequency domain to remove interference from the recording equipment at 50, 28 and 22 Hz, and transform each channel to have zero mean and unit variance.

The multivariate embedding of dimension $m$ is made by using the first $m$ channels of the data. In Fig. 2, we show the Takens estimator of correlation dimension with an upper cutoff $r_o = 0.5$ and $W = 20$ for the first 8192 points of the data set, both for the original data (solid line) and for the amplitude adjusted multivariate surrogates (dots). While there is no indication of low dimensionality, there is evidence for nonlinearity, as the statistic for the original data is significantly less than that for the surrogates, at least for the smaller embedding dimensions. However, the difference between the statistic for the original data, and that for some of the surrogates is only a few percent, so while the difference is significant, it is not substantial.

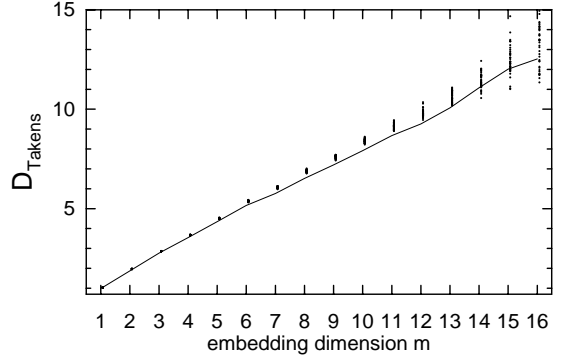

FIG. 2. Takens dimension estimator with $r_o = 0.5$ as a function of embedding dimension, for multivariate embeddings using the first 1 through 16 channels of the EEG data. Solid curve is for the original data, dots are for the amplitude adjusted multivariate surrogates.

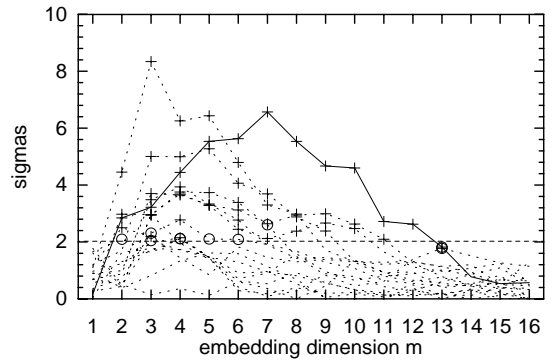

FIG. 3. Significance (measured in sigmas) of nonlinearity for multivariate (solid) and univariate (dotted) embeddings as a function of embedding dimension for the first 8192 points. Pluses indicate points for which the value of the Takens estimator for the original data is outside the distribution of values for the 39 surrogates; this corresponds to a rejection of the null hypothesis at the 95% confidence level. The dashed line is the approximate 95% confidence limit based on a $t$-distribution with 38 degrees of freedom. Circles are for discrepancies between the exact and approximate 95% confidence limits. The confidence limits derived from the $t$-distribution are in surprisingly good agreement with the bootstrapped values.

We also consider each of the 16 channels individually; here, we use a time delay embedding with a lag of $\tau = 2$ sample times (the point at which the autocorrelation function was roughly one half). 39 univariate amplitude adjusted surrogate data sets are generated for each of the channels. The Takens estimator is then computed, again using $r_o = 0.5$ and $W = 20$, for the original and surrogates of each channel, using embedding dimensions of 1 through 16. As above, we use the number of "sigmas" as a rough measure of significance, and in Fig. 3 we show the results for both the multivariate (solid lines) and univariate (dotted lines) data sets. For smaller embedding dimensions, channels 2 and 13 — corresponding to the right occipital (O2) and left frontal (F7) sites — give the best evidence for nonlinearity (largest sigma values), while for larger embedding dimensions the multivariate embedding is better. We remark that the full range of embedding possibilities has not been considered; in particular, we suspect that the "optimal" embedding will be a combination of some channels and some time delays.

## ACKNOWLEDGMENTS

We are grateful to Milan Paluš for providing the EEG data, and for his comments and suggestions. We also thank Danny Kaplan and Paul Rapp for many useful discussions. DP is partially supported by NSF grant ATM-9213522. JT is partially supported by NIMH grant 1-R01-MH47184 and by the US Department of Energy.